\documentclass[aps,10pt,pre,floatfix,twocolumn]{revtex4-1}
\usepackage{ulem}
\usepackage{color}
\usepackage{graphicx}
\usepackage{subfigure,amsmath, amsthm, amssymb}
\usepackage[colorlinks,citecolor=red]{hyperref}
\newcommand{\be}{\begin{equation}}
\newcommand{\ee}{\end{equation}}
\newcommand{\bea}{\begin{eqnarray}}
\newcommand{\eea}{\end{eqnarray}}

\begin{document}
\title{The columnar-disorder phase boundary   in a mixture of hard squares and dimers}
\author{Dipanjan Mandal}
\email{mdipanjan@imsc.res.in}
\affiliation{The Institute of Mathematical Sciences, C.I.T. Campus,
Taramani, Chennai 600113, India}
\affiliation{Homi Bhabha National Institute, Training School Complex, Anushakti Nagar, Mumbai 400094, India}
\author{R. Rajesh} 
\email{rrajesh@imsc.res.in}
\affiliation{The Institute of Mathematical Sciences, C.I.T. Campus,
Taramani, Chennai 600113, India}
\affiliation{Homi Bhabha National Institute, Training School Complex, Anushakti Nagar, Mumbai 400094, India}

\date{\today}
\begin{abstract}
A mixture of hard squares, dimers and vacancies on a square lattice 
is known to undergo a transition from a low-density disordered phase to high-density columnar ordered phase. 
Along the fully packed square-dimer line, the system
undergoes an Kosterliz-Thouless type transition to a phase with power law correlations.
We estimate the phase boundary separating the ordered and disordered phases by calculating the interfacial tension 
between two differently ordered phases within two different approximation schemes.
The analytically obtained phase boundary is in good agreement with Monte Carlo simulations.
\end{abstract}

\maketitle
\section{Introduction}

It is well-known that the universal features of continuous phase transitions may be determined
by studying simple model systems. Among such models, hard core lattice gases (HCLGs) 
are one of the simplest. In these lattice models, particles interact only through excluded
volume interactions and phase transitions, if any, depend only on the shape of the
constituent particles and do not depend on temperature. Hence, they have often been termed
as geometrical or entropy-driven transitions. Despite is simplicity, HCLGs exhibit 
different broken-symmetry phases like
solid-like sublattice order, columnar or smectic phase with partial translational order, and
nematic phase with orientational order, depending on the particle shape. Many monodispersed systems
of differently shaped particles have been studied in the literature. 
Examples include triangles~\cite{verberkmoes}, squares~\cite{bellerman,bellerman2,
francis,kabir1,trisha2,dipanjan}, Y-shaped molecules~\cite{ruth}, 
dimers~\cite{kasteleyn,temperley,huse,nicholls}, trimers~\cite{ghosh2007random}, tetrominoes~\cite{linyong,barnes},
rods~\cite{deepak1,joyjit1,oettel,vigneshwarn,oettel2}, 
rectangles~\cite{joyjit2,joyjit3,trisha3,gurin}, discs~\cite{fernandez,trisha1}, and hexagons~\cite{baxter1}, the 
latter being the only exactly solvable case.  Polydispersity can hardly be avoided in 
experiments. However, compared to the monodispersed systems, the phenomenology of HCLG models of mixtures of particles
of different shapes is less understood.

Amongst mixtures, the best studied example is that of depletion interaction in mixtures
of particles with small excluded volume and particles with larger excluded volume. When the
excluded volumes are on-site and first nearest neighbour, then  in two
dimensions, it is known from different numerical studies that there is a critical line
ending in a tricritical point separating a low-density  disordered fluid-like phase from a high-density 
solid-like sublattice phase~\cite{stilck2,poland,jiang,jeroen,alain,frenkel}. The nature of the transition is similar
to that of the transition observed in the system with only larger particles.
Similar demixing transition occurs in binary mixture of large and small
cubes~\cite{dijkstra1,dijkstra2,lafuente}.
Other examples of 
mixtures that have been studied on lattices  include bidispersed rods~\cite{sr15,ksr16} in which the phase diagram 
is richer than the monodispersed case, but without the appearance of any new phase.  
In a recent paper~\cite{kabir2}, a mixture of hard  squares and dimers was studied and, 
very interestingly, it was demonstrated that the phase diagram consists of
two critical lines with continuously varying exponents that meet at a point called as the Askin-Teller-Kosterlitz-Thouless point. 
It provided the first example of a HCLG whose critical properties vary with the composition. In this
paper, we focus on aspects of this mixture of dimers and squares.

We first discuss the known results for monodispersed systems of dimers or squares. The dimer
model~\cite{kasteleyn,temperley,fisher1,fisher2,lieb,huse,heilmann}, in which each particle occupies 
$1\times 2$ or $2\times 1$ sites on the square lattice
is the simplest model for anisotropic particles. It is exactly solvable in the fully packed limit. In this
limit, the system is critical and the correlations between dimers decay with separation $r$ as
as $r^{-2}$~\cite{fisher2}. At densities different from the fully packed limit, the system is disordered~\cite{lieb,heilmann}.

The hard square model~\cite{bellerman,bellerman2,francis,kabir1,dipanjan} in which each particle 
occupies  $2\times 2$ sites on the square
lattice is the prototypical model to show columnar order. As density is increased, the
system undergoes a continuous phase transition from a low-density
disordered phase to a high density columnar ordered 
phase with a four-fold symmetry. In the columnar phase, the squares preferentially occupy 
either even or odd rows or even or odd columns, thus breaking translational order in
only one of the two directions. The continuous transition belongs to the Ashkin-Teller universality
class with the correlation length exponent $\nu\approx 0.92$~\cite{zhitomirsky,feng,kabir2}.

The mixture of hard squares and dimers was studied both numerically and analytically in Ref.~\cite{kabir2}.
The system undergoes a transition from a square-rich columnar phase
to a dimer-rich disordered phase across a critical line along which the critical exponents continuously
vary depending on the composition of the mixture, consistent with  the Ashkin-Teller universality class. 
On the fully packed line, it was shown that the
system undergoes a Korterlitz-Thouless type transition from a columnar phase to a power law correlated
phase as the dimer density is increased. Along the fully packed line, the configurations of dimers and squares may
be mapped onto a height field. By writing an effective Hamiltonian for the two dimensional height
field, it was possible to theoretically explain the numerically obtained results, along the fully packed 
line~\cite{kabir2}. However, the height mapping does not allow the phase boundary to be determined, as the
relation between the rigidity and microscopic parameters is not easy to establish.
In this paper, we determine the phase diagram within an approximation scheme, and compare with the numerically
obtained phase boundary.

Estimates of phase boundaries in systems showing columnar order, obtained  from standard approximation schemes
like density functional theory, high density expansions, Flory-type approximations, etc., are quite poor [see
Ref.~\cite{dipanjan} for a tabulation of results for the hard square model]. In recent work~\cite{trisha2,dipanjan},
we described a systematic way of determining the interfacial
tension between two differently ordered columnar phases in terms of number of defects and overhangs in the
interface. The estimates obtained from the interfacial tension
are in good agreement with the numerical results for the hard square gas~\cite{trisha2,dipanjan} as well as the hard rectangle
gas~\cite{trisha2}. In this paper, we use the same method to obtain
the phase boundary for the mixture of squares and dimers.
We estimate the interfacial tension between two 
different columnar ordered phases and by setting it to zero, we obtain limiting condition
for the stability of columnar phase.  First, we assume that  the interface between the two ordered phases has no overhangs 
and that the ordered phases have perfect order.
We improve the estimate for the phase boundary by allowing the interface to have overhangs of 
height one. The results are summarized in Fig.~\ref{sublat_fig}. For instance, along the fully packed 
line our estimates for the critical density for squares are within $8 \%$ of the numerical result.

The remainder of the paper is organized as follows. In section~\ref{sec:2}, we define  the 
model precisely and give an outline of  the calculation. The details of the calculations are 
presented in Sec.~\ref{sec:3} and Sec.~\ref{sec:4}, and the results are compared with
results from Monte Carlo simulations.
We end  with a discussion in Sec.~\ref{sec:5}.

\section{\label{sec:2}Model and outline of calculation}

Consider a square lattice of size $L_x \times L_y$. The lattice
may be occupied by particles of three different shapes: squares, horizontal dimers and
vertical dimers of size $(2\times2)$, $(2\times1)$ and $(1\times2)$ respectively. 
The particles interact only through excluded volume interaction, i.e. no two particles may overlap. 
We associate with each square, horizontal dimer, vertical dimer and vacancy ($1\times1$)
activities $z_s$, $z_h$, $z_v$ and $z_0$  respectively. We will refer to the bottom
left corner of a particle as its head.

Depending on the values of the activities, the system may exist in a disordered
fluid like phase or in an ordered phase which has columnar order~\cite{kabir2}. 
In the columnar phase,
the heads of squares and vertical (horizontal) dimers 
preferably occupy either even or odd rows (columns) with  equal fraction on an average in
even or odd columns (rows). 
 
\begin{figure}
\includegraphics[width=\columnwidth]{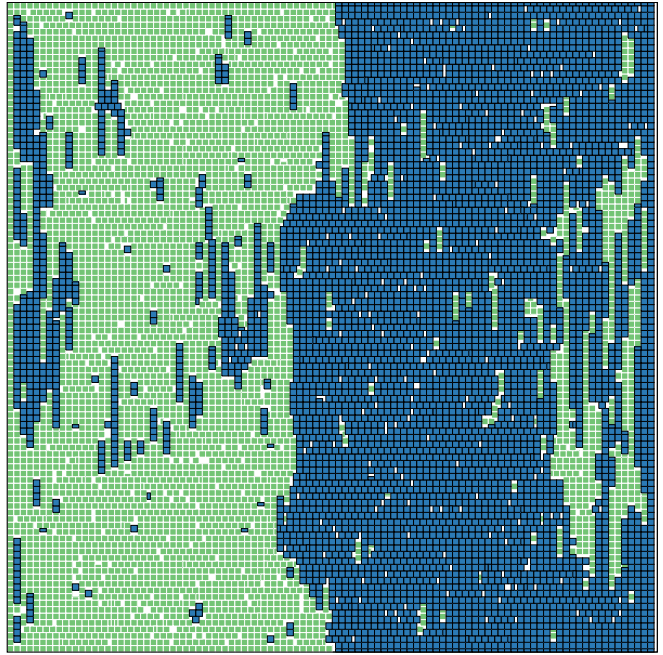}
\caption{Snapshot of a typical equilibrium configuration of a system where the squares at the left boundary
are fixed to be on even rows (green) and the squares at the right boundary are fixed to be on
odd rows (blue). At high enough activity $z_s$ (as in the figure), a sharp interface separates the left even
phase from the right odd phase.
}
\label{model}
\end{figure}

The aim of the paper is to determine the phase
boundary between the columnar and disordered phases. This is done by estimating
the interfacial tension $\sigma(z_s,z_h,z_v,z_0)$ between two ordered columnar phases,
and by equating it to zero, we obtain estimates for the
critical activities and densities. 
Let the phase in which majority of heads of squares and vertical dimers 
are in even (odd) rows be called even (odd) phase.
To compute $\sigma(z_s,z_h,z_v,z_0)$, we impose an interface  in the system
by fixing squares at the left boundary to be even and those at the right boundary to be
odd.  A snapshot of a typical equilibrium  configuration seen in a Monte Carlo simulation of a system with
these boundary conditions is
shown in Fig.~\ref{model}. 
There is a left phase with even squares and vertical dimers separated from a
right phase with odd squares and vertical dimers by an interface. The horizontal dimers
could be even or odd in both phases. To define an unique position 
of the interface for any allowed configuration of particles, we adopt the convention that 
the boundary between the left or even phase and right
or odd phase is placed as far left as possible. With this convention, 
there is a  well-defined interface. The bulk phases have only few defects. By defects,
we mean  particles of the wrong type (odd in even phase or even in odd phase). 
When the defects are removed, a 
fully ordered columnar phase is recovered.

Let $Z^{(0)}$ and $Z^{(I)}$ be the partition function of the 
system without and with an interface $I$ respectively. The interfacial tension $\sigma$ is defined as
\be
\label{sutface tension}
e^{-\sigma L_y}=\frac{\sum_I Z^{(I)}}{Z^{(0)}},
\ee
Due to the nature of interactions between particles being hard core,
we can write the partition function of the system in the presence of 
an interface as a product of the partition functions of the left
and right phases, i.e.
\begin{align}
\label{interface}
 Z^{(I)}=Z_L^{(I)}Z_R^{(I)},
\end{align}
where $L$ and $R$ denote left and right.

\begin{figure}
\includegraphics[width=\columnwidth]{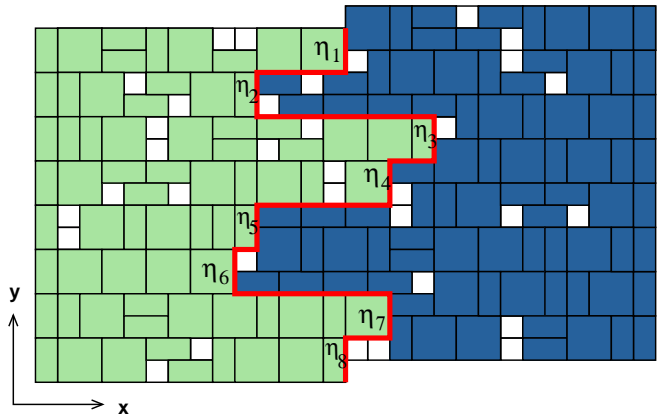}
\caption{ A schematic diagram of an interface that has no overhangs. The interface is indicated  by the
red line and its $x$-coordinates are denoted by $\eta_i$. The boundary conditions are periodic
in the  $y$-direction.}
\label{f4}
\end{figure}

\begin{figure}
\includegraphics[width=0.6\columnwidth]{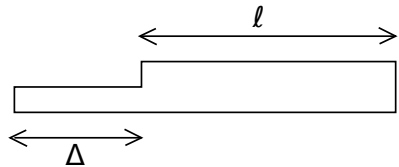}
\caption{The shape of a generic track  of two rows. It is characterized by two lengths
$\ell$ and $\Delta$ and has partition function $\Omega(\ell,\Delta)$.}
\label{delta1}
\end{figure}

$Z^{(I)}$ cannot be calculated for an arbitrary interface $I$. We therefore calculate it
within two approximations. As a
first approximation, we consider the simplest case where we ignore 
overhangs in the interface and defects in the bulk.
In this simplified model, the interface is defined by the 
position of right boundary of the left phase, and denoted by $\eta_i$ (see Fig.~\ref{f4}). Since
we assume perfect columnar order for the left and right phases, the partition functions
for both left and right phases  are a  product of partition
functions of tracks made up of two adjacent rows ($L_y/2$ of them):
\begin{align}
\label{left}
Z_L^{(I)}=&\prod_{i=1}^{L_y/2} \big[z_v \Omega(\eta_i -1,0)+z_s \Omega(\eta_i-2,0)\big],\\
\label{right}
Z_R^{(I)}=&\prod_{i=1}^{L_y/2}\Omega(L_x-\max(\eta_i,\eta_{i+1}),|\eta_i-\eta_{i+1}|),
\end{align}
where $\Omega(\ell,\Delta)$ 
is the partition function of a track of two rows with a shape as shown in Fig.~\ref{delta1}.
In the region corresponding to $\Delta$, only horizontal dimers can be placed.
The right hand side of Eq.~\eqref{left} follows from the fact that, for the left phase,
there must be either a 
vertical dimer or square touching the interface.
The partition function of the system without an interface is
\begin{align}
\label{no_interface}
 Z^{(0)}=\prod_{i=1}^{L_y/2}\Omega(L_x,0).
\end{align}

In the second approximation, we allow the interface to have overhangs of height one. We still do not allow defects
in the bulk. The calculation is on the same lines as that described above.
This allows us to obtain an improved estimate of the critical parameters. 

The calculation of the interfacial tension involves determining the partition function $\Omega(\ell, \Delta)$ of a track
of two rows, and is done in the next section.

\section{\label{sec:3}Calculation of two-row partition function $\Omega(\ell, \Delta)$}

In this section we calculate $\Omega(\ell, \Delta)$, the partition function of a track of two
rows with shape as shown in Fig.~\ref{delta1}. Consider the generating function
defined as
\begin{equation}
\label{eq gy1}
G(y,\Delta)=\sum_{\ell=0}^{\infty}\Omega(\ell,\Delta)y^{\ell+\Delta/2}, 
\end{equation}
where the power of $\sqrt{y}$ is the number of sites in the system.

\begin{figure}
\includegraphics[width=\columnwidth]{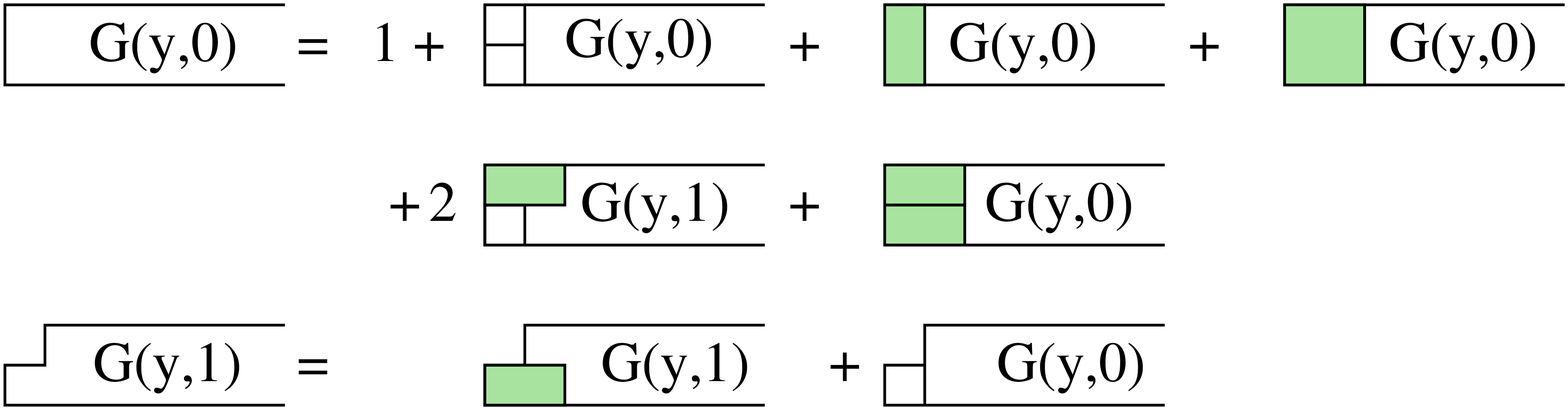}
\caption{Diagrammatic representation of the recursion relations 
obeyed by the generating functions $G(y,0)$ and $G(y,1)$ [see Eq.~\eqref{eq gy1} for definition]. The left-most column
may be occupied by  vacancies, dimers or squares.}
\label{two_row_gf}
\end{figure}

First, consider the case $\Delta=0,1$. $G(y,0)$ and $G(y,1)$ obey simple recursion 
relations which are shown diagrammatically in Fig.~\ref{two_row_gf} and may be 
written as 
\begin{eqnarray}
\label{eq gya}
G(y,0)&=&1 + z_0^2 y G(y,0) + z_v y G(y,0)+z_s y^2 G(y,0) +\nonumber \\ 
&&2 z_0z_h y^{3/2} G(y,1)+ z_h^2 y^2 G(y,0),\\
G(y,1)&=&z_h y G(y,1)+z_0 y^{1/2} G(y,0).
\label{eq gyb}
\end{eqnarray}
These are easily solved to yield
\begin{align}
\label{generating_fn0}
G(y,0)&=\frac{1-z_h y}{f(y)},\\
G(y,1)&=\frac{z_0 \sqrt{y}}{f(y)},
\label{generating_fn1}
\end{align}
where the function $f(y)$ in the denominator is
\begin{align}
f(y)=&z_h(z_s+z_h^2)y^3-(z_s+z_h^2+z_hz_0^2-z_h z_v)y^2\nonumber\\
&-(z_h+z_v+z_0^2)y+1,
\end{align}
a third order polynomial in $y$.
Let $y_1$ be smallest root of $f(y)=0$. Then, it is clear that
\be
\label{approximated}
\Omega(\ell,\Delta)=a(\Delta) \lambda ^\ell [1+ O(e^{-c \ell})], ~~\ell\gg 1,
\ee
where 
\begin{equation}
 \lambda=\frac{1}{y_1}.
\end{equation}
The prefactor $a(\Delta)$ for $\Delta=0,1$ is determined by calculating the
coefficient of $y^{\ell+\Delta/2}$. It is easily checked that
\begin{align}
a(0)=&\frac{-(1-z_h y_1)}{y_1f^\prime(y_1)}, \label{eq:a0}\\
a(1)=&\frac{-z_0}{y_1f^\prime(y_1)}. \label{eq:a1}
\end{align}
Equations~\eqref{approximated}--\eqref{eq:a1} determine
$\Omega(\ell, \Delta)$ for $\Delta=0,1$.

\begin{figure}
\includegraphics[width=\columnwidth]{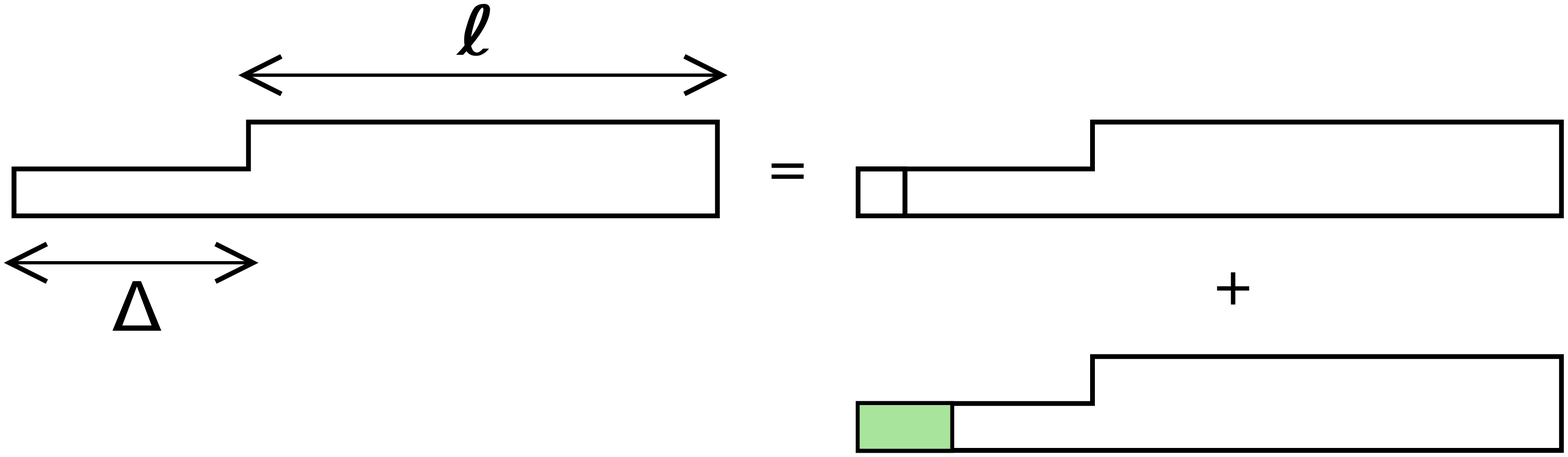}
\caption{ Diagrammatic representation of the recursion relation obeyed by
$\Omega(\ell,\Delta)$, the partition function of a track of two rows as shown in Fig.~\ref{delta1}, for $\Delta \geq 2$. 
The left-most column of the 
may be occupied by a vacancy or a horizontal dimer.}
\label{fig h}
\end{figure}

We now calculate $\Omega(\ell,\Delta)$ for $\Delta\geq2$. The recursion relation 
obeyed by $\Omega(\ell,\Delta)$ is shown diagrammatically
in Fig.~\ref{fig h}, and may be written as
\be
\label{genfn_H}
\Omega(\ell,\Delta)=z_0 \Omega(\ell,\Delta-1)+z_h \Omega(\ell,\Delta-2),~\Delta\geq 2.
\ee
To solve Eq.~\eqref{genfn_H}, define the generating function
\be
H(\ell,x)=\sum_{\Delta=0}^\infty \Omega(\ell,\Delta)x^\Delta.
\ee
Multiplying Eq.~(\ref{genfn_H}) by $x^\Delta$ and summing over $\Delta$, we obtain
a linear equation for $H(\ell,x)$ that may be  solved to yield
\be
H(\ell,x)=\frac{\Omega(\ell,1)x+\Omega(\ell,0)(1-z_0x)}{1-z_0x-z_hx^2}.
\label{Hlx}
\ee

The generating function $H(\ell,x)$ has two simple poles determined by the
zeros of the quadratic equation $1-z_0x-z_h x^2=0$. These are
\begin{align}
x_{\pm}=\frac{-(z_0\pm \sqrt{z_0^2+4z_h})}{2z_h}.
\label{xpm}
\end{align}
Expanding the denominator of Eq.~(\ref{Hlx}) about $x_+$ and $x_-$ we obtain
the coefficient of $x^\Delta$ to be 
\begin{align}
\Omega(\ell,\Delta)=\sum_{i=\pm}\frac{a(1)+a(0)z_h x_i}{(z_0+2z_hx_i)x_i^\Delta}
\lambda^{\ell},\ell \gg 1.
\label{omegauv}
\end{align}
Using Eqs.~(\ref{approximated}) and (\ref{omegauv}), we can write
\begin{align}
\label{a_delta}
a(\Delta)=\frac{p_+}{x_+^\Delta}+\frac{p_-}{x_-^\Delta},\Delta\geq0,
\end{align}
where 
\begin{align}
p_\pm=\frac{a(1)+a(0)z_hx_\pm}{z_0+2z_hx_\pm}.\nonumber
\end{align}

\section{\label{sec:4}Interfacial tension and critical parameters}

We now calculate  interfacial tension $\sigma(z_s,z_h,z_v,z_0)$.
The results for the interface without overhangs is in Sec.~\ref{sec:no_overhang}
and for the interface with overhangs in Sec.~\ref{sec:overhang}.

\subsection{Without overhangs\label{sec:no_overhang}}

For large $L_x$ and $\eta_i$, using Eqs.~(\ref{left}), (\ref{right}) and (\ref{approximated}), the
partition functions of the left and right domains may be written as
\bea
\label{il}
Z_L^{(I)}&=&\bigg[\frac{a(0)}{\lambda}\big(z_v+\frac{z_s}{\lambda}\big)\bigg]^{L_y/2}
\prod_{i=1}^{L_y/2}\lambda^{\eta_i} ,\\
\label{ir}
Z_R^{(I)}&=&\prod_{i=1}^{L_y/2}a(|\eta_i-\eta_{i+1}|)\lambda^{L_x-\max(\eta_i,\eta_{i+1})}.
\eea
Taking the product of Eqs.~(\ref{il}) and (\ref{ir}) and using the relation
\be
\max(\eta_i,\eta_{i+1})=\frac{|\eta_i-\eta_{i+1}|+\eta_i+\eta_{i+1}}{2},
\ee
we obtain the partition function of the system with interface
\bea
\label{z_interface}
Z^{(I)}&=&\bigg[\frac{a(0)}{\lambda}(z_v+\frac{z_s}{\lambda})\bigg]^{L_y/2}\times\nonumber\\
&\prod&_{i=1}^{L_y/2}\bigg[\lambda^{L_x-|\eta_i-\eta_{i+1}|/2}a(|\eta_i-\eta_{i+1}|)\bigg].
\eea
Note that $Z^{(I)}$ depends only on the difference $|\eta_i-\eta_{i+1}|$ and not on $\eta_i$. 
Summing over all configurations $I$ is equivalent to summing  over all differences $(\eta_i-\eta_{i+1})$.
Performing the sum and using Eq.~(\ref{a_delta}), we obtain
\begin{align}
\sum_I Z^{(I)}&=\Bigg[a(0)\lambda^{L_x-1}(z_v+\frac{z_s}{\lambda})\times \nonumber\\
&\bigg( p_+\frac{x_+\sqrt{\lambda}+1}{x_+\sqrt{\lambda}-1}+
p_- \frac{x_-\sqrt{\lambda}+1}{x_-\sqrt{\lambda}-1}\bigg)\Bigg]^{L_y/2}.
\label{eq:26}
\end{align}

The partition function of the system without an interface can be easily calculated using
Eq.~(\ref{no_interface}) 
\be
\label{z_no_interface}
Z^{(0)}=\bigg[\Omega(L_x,0)\bigg]^{L_y/2}=\bigg[a(0) \lambda^{L_x}\bigg]^{L_y/2}.
\ee
The interfacial tension, as defined in Eq.~\eqref{sutface tension} may be determined from 
Eqs.~\eqref{eq:26} and \eqref{z_no_interface} to be
\be
e^{-\sigma L_y}=\Bigg[(\frac{z_v}{\lambda}+\frac{z_s}{\lambda^2}) 
\bigg( p_+\frac{x_+\sqrt{\lambda}+1}{x_+\sqrt{\lambda}-1}+
p_- \frac{x_-\sqrt{\lambda}+1}{x_-\sqrt{\lambda}-1}\bigg)\Bigg]^{\frac{L_y}{2}}.
\ee
The phase boundary corresponds to the values of the parameters at
which the interfacial tension vanishes. This immediately gives
\be
\label{sigma zero}
\left[\frac{z_v}{\lambda}+\frac{z_s}{\lambda^2}\right]\left[
p_+\frac{x_+\sqrt{\lambda}+1}{x_+\sqrt{\lambda}-1}+p_- 
\frac{x_-\sqrt{\lambda}+1}{x_-\sqrt{\lambda}-1}\right]=1.
\ee

\begin{figure}
\subfigure[]{\includegraphics[width=\columnwidth]{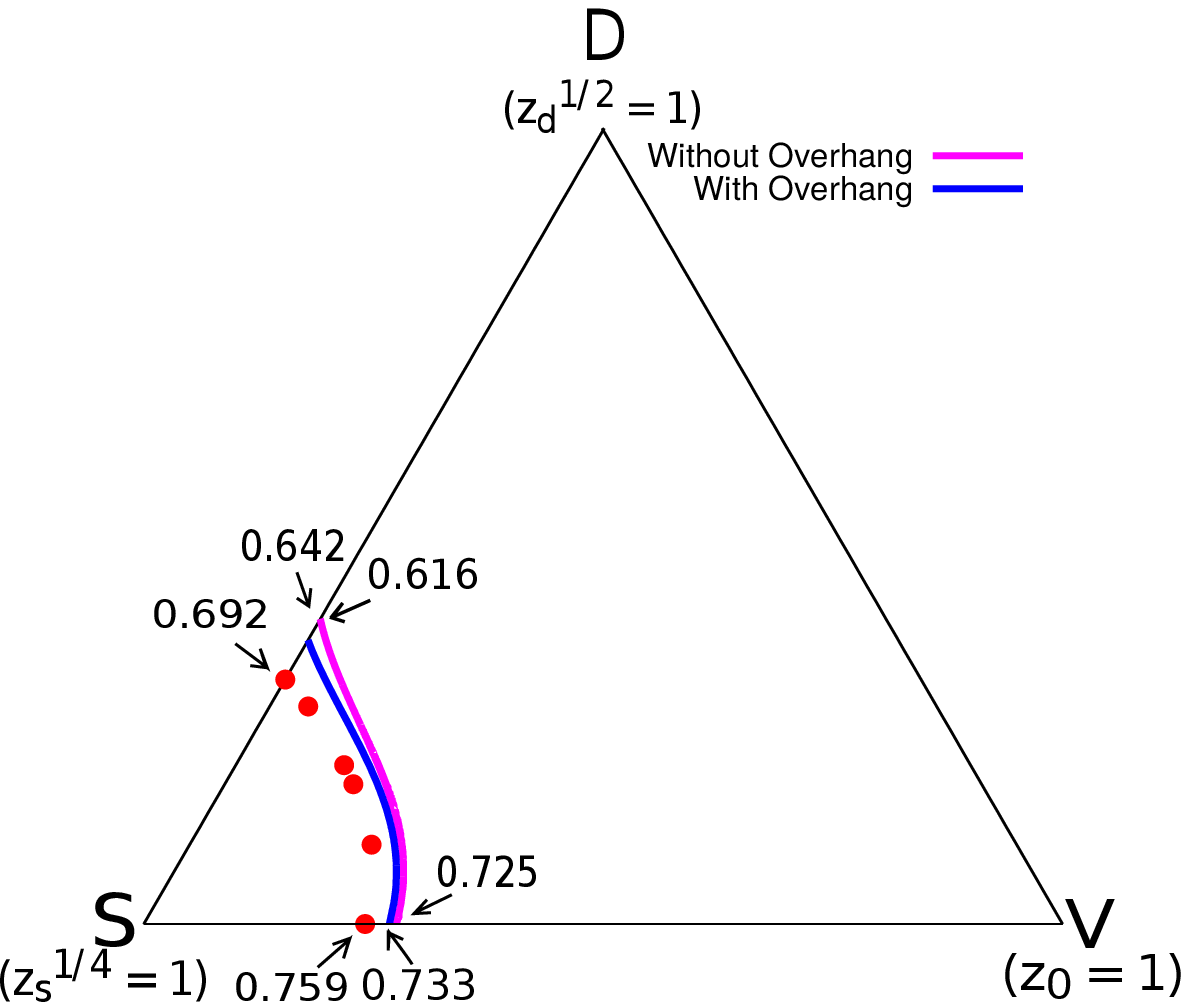}\label{f6}}
\subfigure[]{\includegraphics[width=\columnwidth]{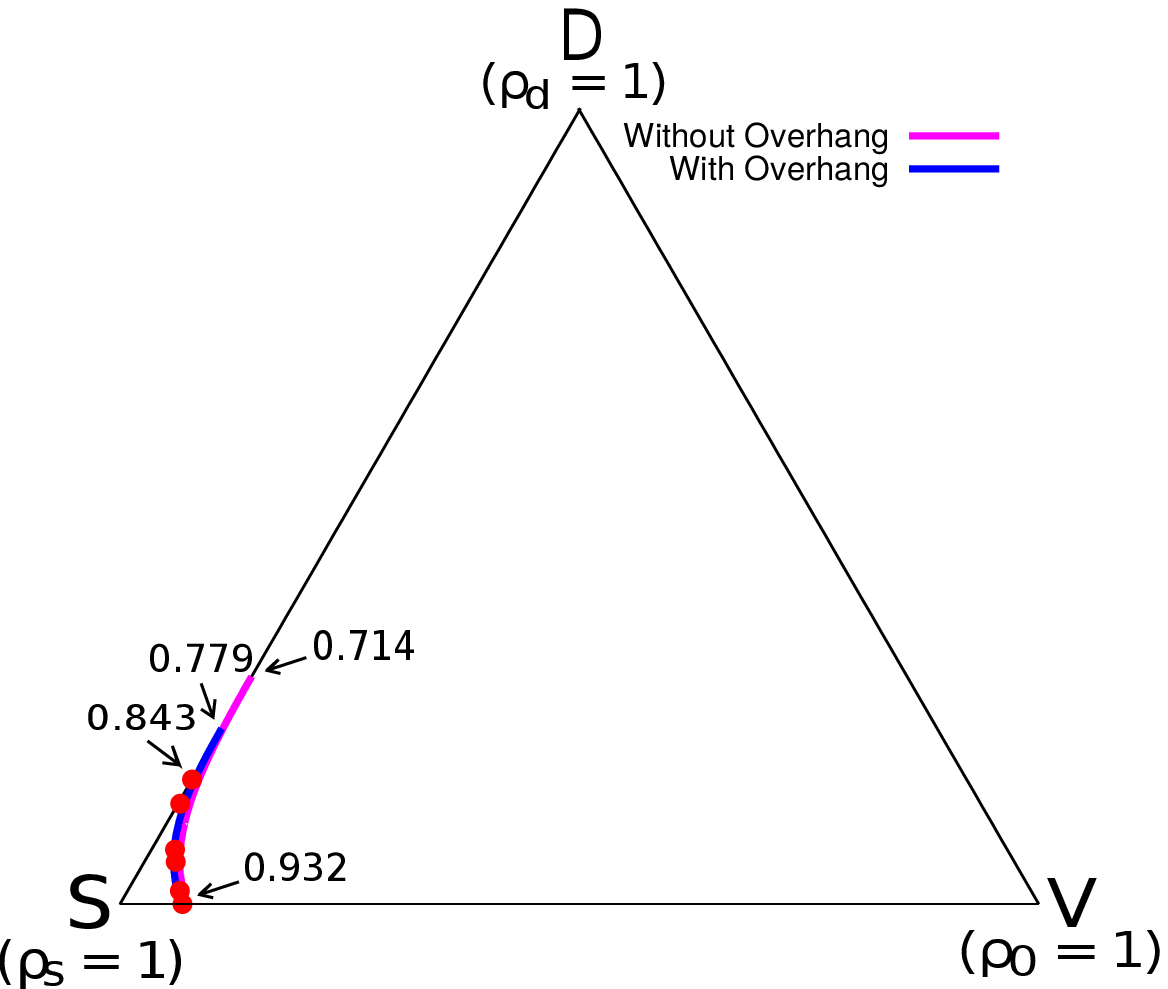}\label{f7}}
\caption{ Phase diagram  in 
(a) activity z-plane and (b) density $\rho$-plane. $S$ and $D$  represents the state where the 
lattice is fully packed  by squares and dimers respectively. $V$ represents the empty lattice.
The estimates of the phase boundaries obtained from modelling the interface without overhangs is
shown by magenta line while that obtained by including overhangs of height one
are shown by blue lines. The data points (red circles) are obtained from Monte Carlo simulations (see Sec.~\ref{sec:numerical}).
}
\label{sublat_fig}
\end{figure}

The phase boundary obtained from Eq.~\eqref{sigma zero} is shown by the 
magenta line in Fig.~\ref{f6} and Fig.~\ref{f7}. Here, the activities have been normalized
using
\be
\label{zsdv}
z_s^{1/4}+z_d^{1/2}+z_0=1,
\ee
where $z_d=z_h=z_v$, such that a two dimensional phase diagram may be obtained.
It shows transitions at $z_s^{1/4}=0.725$ along square-vacancy (SV) line and at
$z_s^{1/4}=0.616$ along square-dimer (SD) line. 
These should be compared with results from Monte Carlo simulations~\cite{kabir2}: 
$z_s^{1/4}=0.759$ along SV line and at $z_s^{1/4}=0.692$ along SD line. 

The density of sites occupied by squares $\rho_s$, horizontal dimers $\rho_{h}$ and vertical dimers $\rho_{v}$ 
may be calculated from the partition function $Z^{(0)}$ as:
\bea
\rho_s&=&\frac{4 z_s}{L_xL_y}\frac{\partial \ln Z^{(0)} }{\partial z_s},\\
\rho_i&=&\frac{2z_i}{L_xL_y}\frac{\partial \ln Z^{(0)}}{\partial z_i},~i=h,v,
\eea
where the factor 4 and 2 accounts for the area of a square and a dimer respectively.
Substituting for $Z^{(0)}$ from Eq.~(\ref{z_no_interface}), the densities may be written in the thermodynamic
limit $L_x \rightarrow \infty$, $L_y \rightarrow \infty$ as:
\bea
\rho_s&=&\frac{2z_s}{\lambda}\frac{\partial\lambda}{\partial z_s},\\
\rho_i&=&\frac{z_i}{\lambda}\frac{\partial\lambda}{\partial z_h},~i=h,v,
\eea
where total dimer density $\rho_d=\rho_h+\rho_v$ and density of vacancy $\rho_0=1-\rho_d-\rho_s$.

In the density plane, it shows transitions at 
$\rho_s=0.928$ along SV line and at
$\rho_s=0.714$ along SD line. The Monte Carlo results show transitions at
$\rho_s=0.932$ along SV line and at
$\rho_s=0.843$ along SD line.
Our estimation of critical activities along SV and SD lines agree satisfactorily with the numerical results
(see Sec. \ref{sec:numerical}).

\subsection{With overhangs \label{sec:overhang}}

In the calculation presented in Sec. \ref{sec:no_overhang}, the interface was modelled as 
having no overhangs.
We now allow the interface to have a certain subclass of overhangs, and obtain an improved estimate for the
phase boundary. To be able to do the calculation in the presence of overhangs, we first reformulate the calculation
in the absence of overhangs in terms of a weighted, directed  walk.

The interface with no overhang (see Fig.~\ref{f4}) may be  visualized  as a partially directed self avoiding walk (PDSAW) 
from top to bottom, where the walk is not allowed to take a step in the upward direction,  but allowed to take steps in the leftward,
rightward, and downward directions as long as the walk is self avoiding. We denote a downward step
by $\mathbb{D}$, a leftward step of  length $\Delta$ by $\mathbb{L}_\Delta$ and rightward step of  length $\Delta$ by 
$\mathbb{R}_\Delta$. To maintain self avoidance, every set of consecutive  leftward or rightward steps must be preceded (or followed) by
a downward step. Thus, the different PDSAWs may be enumerated by arbitrary concatenation of substrings $\mathbb{D}$, 
$\mathbb{D} \mathbb{L}_\Delta$, and $\mathbb{D} \mathbb{R}_\Delta$ where $\Delta=1,2,\ldots$, and the length of the walk in the vertical
direction is given by the number of $\mathbb{D}$s in the string. The formal generating function of these strings may be written as
\bea
\mathbb{G}_0&=&\sum_{L_y=0,2,4,\ldots}(\mathbb{D}+\mathbb{D}\widetilde{\mathbb{R}}+\mathbb{D}\widetilde{\mathbb{L}})^{L_y/2},\label{eq:g0}\\
&=&\frac{1}{1-(\mathbb{D}+\mathbb{D}\widetilde{\mathbb{R}}+\mathbb{D}\widetilde{\mathbb{L}})},
\eea
where 
\bea
\widetilde{\mathbb{L}}&=&\sum^\infty_{\Delta=1} \mathbb{L}_\Delta,\\
\widetilde{\mathbb{R}}&=&\sum^\infty_{\Delta=1} \mathbb{R}_\Delta.
\eea
The generating function $\mathbb{G}_0$ generates all possible walks of all possible lengths along the vertical direction. 
However, it does not assign weights to a walk.

To assign weights to each walk, we have to
determine the weights $D$, $L_\Delta$ and $R_\Delta$ that correspond to the substrings $\mathbb{D}$, $\mathbb{L}_\Delta$, and 
$\mathbb{R}_\Delta$. To do so, we determine the weight of a  PDSAW by taking the ratio of  Eq.~(\ref{z_interface}) and Eq.~(\ref{z_no_interface}) to obtain
\bea
\frac{Z^{(I)}}{Z^{(0)}}&=&\bigg[\frac{a(0)}{\lambda}\bigg(z_v+\frac{z_s}{\lambda}\bigg)\bigg]^{L_y/2}\times\nonumber\\
&\prod&_{i=1}^{L_y/2}\bigg[\lambda^{-|\eta_i-\eta_{i+1}|/2}\frac{a(|\eta_i-\eta_{i+1}|)}{a(0)}\bigg], \label{weight_walk}\\
&=& D^{L_y/2} \prod_{i=1}^{L_y/2} (R,L)_{|\eta_i-\eta_{i+1}|}, \label{eq:correspondence}
\eea
where $R$ or $L$ in Eq.~(\ref{eq:correspondence}) is chosen depending on whether the step is in the rightward or leftward direction.
From Eqs.~(\ref{weight_walk}) and (\ref{eq:correspondence}), we immediately read out
\bea
\label{down_weight}
 D&=&a(0)\left(\frac{z_s}{\lambda^2}+\frac{z_v}{\lambda} \right),\\
\label{left_weight}
L_{\Delta}&=&\frac{a(\Delta)}{a(0)}\lambda^{-\Delta/2},\\
\label{right_weight}
R_{\Delta}&=&\frac{a(\Delta)}{a(0)}\lambda^{-\Delta/2}.
\eea

Consider now the weighted generating function
\be
\label{eq:xyz}
\mathcal{G}_0=\frac{1}{1-(D+D\widetilde{R}+D\widetilde{L})},
\ee
where
\bea
\label{right_left_wt}
\widetilde{R}&=&\sum^\infty_{\Delta=1}R_\Delta=\sum^\infty_{\Delta=1}L_\Delta=\widetilde{L},\nonumber\\
&=&\frac{1}{a(0)}\bigg(\frac{p_+}{x_+\sqrt{\lambda}-1}+\frac{p_-}{x_-\sqrt{\lambda}-1}\bigg).
\eea
It is straightforward to see that, in terms of the interfacial tension $\sigma$, $\mathcal{G}_0$ may be written as
\be
\label{eq:G}
\mathcal{G}_0=\sum_{L_y=0}^{\infty}e^{-2 \sigma L_y},
\ee
Equation~(\ref{eq:G}) is convergent for all $\sigma>0$, and diverges at $\sigma=0$, corresponding to the transition point.
From Eq.~(\ref{eq:xyz}), the divergence of $\mathcal{G}_0$ corresponds to  
the condition
\be
\label{nd_c}
D+D\widetilde{R}+D\widetilde{L}=1,
\ee
Substituting for $D$, $\widetilde{R}$ and $\widetilde{L}$ from Eqs.~(\ref{down_weight}) and (\ref{right_left_wt}), we obtain the same  equation for the phase boundary as obtained earlier in Eq.~(\ref{sigma zero}).

\begin{figure}
\includegraphics[width=\columnwidth]{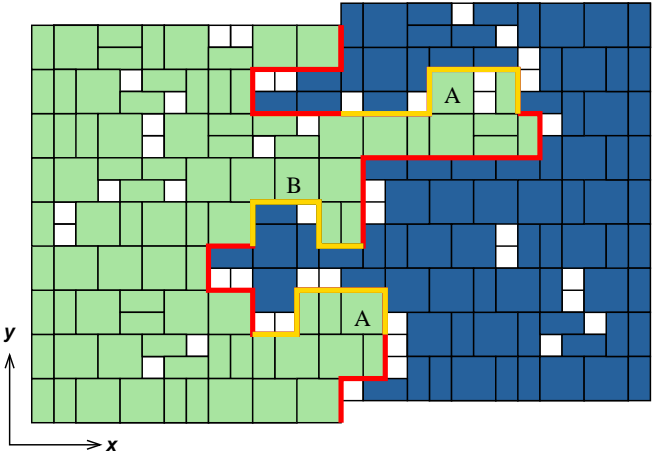}
\caption{ Schematic diagram of an interface with overhangs.  Overhangs are indicated by yellow 
lines while the rest of the interface is indicated by red lines. Overhangs could be of type right (shown by A) or
left (shown by B). The boundary conditions are periodic
in the  $y$-direction.}
\label{f5}
\end{figure}

We now allow the interface to have overhangs of height one. If an overhang is present, then a
horizontal line on the dual lattice will intersect the interface more than once. A schematic diagram  of
an interface with overhangs is shown in  Fig.~\ref{f5}.
The interface with overhangs can no longer be interpreted as a PDSAW, as the walk is now allowed
to take upward steps. However, restricting the overhangs to height one implies that
each upward step has to followed by a downward step before
another upward step can be taken. We now generalize the PDSAW to take into account these
upward steps. 

An overhang will be termed as right or left overhang depending on whether it appears in the right 
(as shown by yellow with index A in Fig.~\ref{f5}) or left phase (as shown by yellow with index B in Fig.~\ref{f5}).
We now separately determine the generating function of all possible walks with right and left overhangs of height one.

\subsubsection{Right Overhangs}

\begin{figure}
\includegraphics[width=\columnwidth]{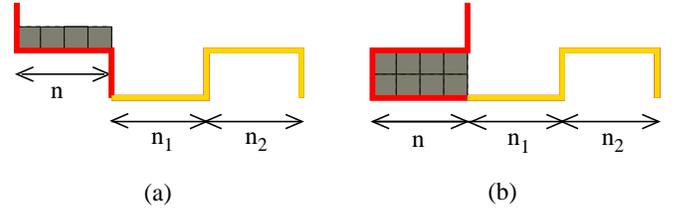}
\caption{ The two kinds of right overhang in which the first downward step is followed by  (a) rightward or (b) leftward step. 
Overhangs are indicated by yellow lines.}
\label{f11}
\end{figure}

There are two kinds of right overhangs depending on whether the first downward step is followed by rightward steps
[see Fig.~\ref{f11}(a)] or leftward steps [see Fig.~\ref{f11}(b)]. We split each of these into two parts: the initial
part shown by red and the remaining part shown by yellow in Fig.~\ref{f11}. This amounts to restricting to
configurations where a horizontal dimer does not cross the right boundary of shaded region in Fig.~\ref{f11}. We 
denote these parts of the walk by $\mathbb{W}_R^{(1)}(n)$, $\mathbb{W}_R^{(2)}(n)$ and $\mathbb{O}_R(n_1,n_2)$ respectively, where the 
superscript refer to the two types of right overhangs and subscript $R$ stands for right. Thus a generic right overhang 
is represented by
\be
\mathbb{W}_R^{(i)}(n)\mathbb{O}_R(n_1,n_2)\mathbb{O}_R(n_3,n_4)...,~i=1,2.
\ee
We now determine the  weights for these overhangs.
It is clear that
\bea
W_R^{(1)}&=&\sum_{n=1}^\infty wt[\mathbb{W}_R^{(1)}(n)]=D^2\sum_{n=1}^\infty\omega(n)\lambda^{-n/2},\\
W_R^{(2)}&=&\sum_{n=0}^\infty wt[\mathbb{W}_R^{(2)}(n)]=D^2\sum_{n=0}^\infty\bigg[\omega(n)\lambda^{-n/2}\bigg]^2,
\eea
where the weight of $\mathbb{W}_R^{(i)}(n)$ may be determined by considering an interface with only one overhang.
Here $\omega(\Delta)$ is the partition function of a track of width one and length $\Delta$, and appears in the weights
because the shaded region in Fig.~\ref{f11} may be occupied by horizontal dimers.

\begin{figure}
\includegraphics[width=\columnwidth]{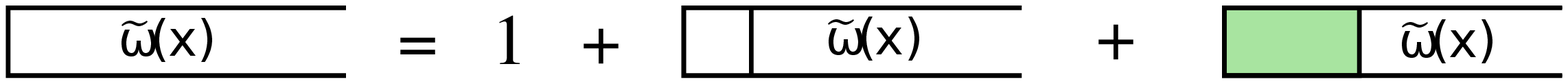}
\caption{Diagrammatic representation of the recursion relation satisfied by the generating function 
$\widetilde{\omega}(x)$
[see Eq.~\eqref{ux} for definition]. The left-most column 
may be occupied by a vacancy or by a horizontal dimer.}
\label{one_row}
\end{figure}

The partition function $\omega(\Delta)$ is easily determined. Define the generating function
\be
\widetilde{\omega}(x)=\sum_{\Delta=0}^\infty \omega(\Delta)x^\Delta,
\label{ux}
\ee
where the power of $x$ represents the number of sites in the system. The recursion relation obeyed by $\widetilde{\omega}(x)$
is shown diagrammatically in Fig.~\ref{one_row} and may be written as
\be
\widetilde{\omega}(x)=1+z_0\widetilde{\omega}(x)x+z_h \widetilde{\omega}(x)x^2,
\ee
with solution
\be
\widetilde{\omega}(x)=\frac{1}{1-z_0x-z_hx^2}.
\ee
Expanding the denominator about its two roots $x_{\pm}$ [see Eq.~(\ref{xpm})] and using Eq.~(\ref{ux}), we 
can write
\be
\label{omega_d}
\omega(\Delta)=\frac{b_+}{x_+^{\Delta}}+\frac{b_-}{x_-^{\Delta}},\Delta\geq0,
\ee
where
\be
b_\pm=\frac{1}{2-z_0x_\pm}.
\label{eq:bplus}
\ee
Using Eq.~(\ref{omega_d}), the weights $W_R^{(1)}$ and $W_R^{(2)}$ may be rewritten as
\be
W_R^{(1)}=D^2\bigg(\frac{b_+}{x_+\sqrt{\lambda}-1}+\frac{b_-}{x_-\sqrt{\lambda}-1}\bigg ),
\ee
\bea
W_R^{(2)}=D^2\bigg(\frac{b_+^2}{1-{(x_+^2\lambda)}^{-1}}+\frac{b_-^2}{1-{(x_-^2\lambda)}^{-1}}\nonumber\\
+\frac{2b_+ b_-}{1-{(x_+x_-\lambda)}^{-1}} \bigg).
\eea

\begin{figure} 
\includegraphics[width=\columnwidth]{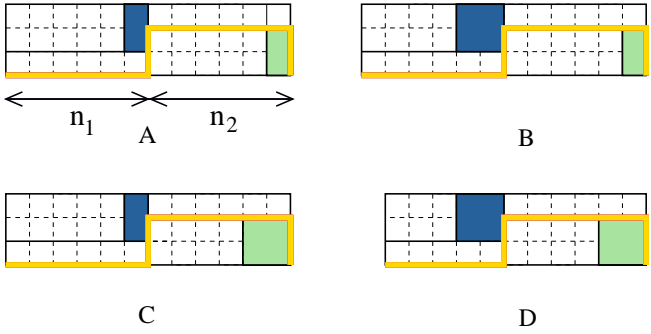}
\caption{ For each right overhang, there are four possible configurations (A--D)   depending on whether 
the particles adjacent to the downward steps are squares or vertical dimers.}
\label{f10}
\end{figure}

Now consider the weight $O_R$ defined as
\bea
O_R&=&\sum_{n_1}\sum_{n_2}wt[\mathbb{O}_R(n_1,n_2)]\nonumber\\
&=&O_R^{(A)}+O_R^{(B)}+O_R^{(C)}+O_R^{(D)},
\eea
where $O_R^{(i)}$ depends on the different ways the particles at the edge of the overhangs may be placed (see Fig.~\ref{f10}). 
The weight of these four kinds of overhangs may be determined in a straightforward manner by considering example of an interface
with only downward step and one overhang :
\bea
\label{ua}
O_R^{(A)}=\sum_{n_1=1}^{\infty}\sum_{n_2=1}^{\infty}\bigg[&z_v^2\omega(n_1)\omega(n_2)
\Omega(n_1-1,0)\nonumber\\ &\Omega(n_2-1,0)\lambda^{-3(n_1+n_2)/2}\bigg],\\
\label{ub}
O_R^{(B)}=\sum_{n_1=2}^{\infty}\sum_{n_2=1}^{\infty}\bigg[&z_vz_s\omega(n_1)\omega(n_2)
\Omega(n_1-1,0)\nonumber\\ &\Omega(n_2-2,0)\lambda^{-3(n_1+n_2)/2}\bigg],\\
\label{uc}
O_R^{(C)}=\sum_{n_1=1}^{\infty}\sum_{n_2=2}^{\infty}\bigg[&z_sz_v\omega(n_1)\omega(n_2)
\Omega(n_1-2,0)\nonumber\\ &\Omega(n_2-1,0)\lambda^{-3(n_1+n_2)/2}\bigg],\\
\label{ud}
O_R^{(D)}=\sum_{n_1=2}^{\infty}\sum_{n_2=2}^{\infty}\bigg[&z_s^2\omega(n_1)\omega(n_2)
\Omega(n_1-2,0)\nonumber\\ &\Omega(n_2-2,0)\lambda^{-3(n_1+n_2)/2}\bigg].
\eea

The sums over $n_i$ may be expressed in term of the generating function $G(y,0)$ [see Eq.~\ref{eq gy1}]. This gives
\bea
O_R&=&\bigg[G(x_+^{-1}\lambda^{-3/2},0)\frac{b_+}{x_+\lambda^{3/2}}\bigg(\frac{z_s}{x_+\lambda^{3/2}}+z_v\bigg)\nonumber\\
&+&G(x_-^{-1}\lambda^{-3/2},0)\frac{b_-}{x_-\lambda^{3/2}}\bigg(\frac{z_s}{x_-\lambda^{3/2}}+z_v\bigg)\bigg]^2.
\label{over_r}
\eea
The total weight of all walks with right overhang  may now be computed. Let $\mathbb{U}_R$ represent all possible walks with right 
overhang and having total weight $U_R$.
Then
\bea
U_R=[W_R^{(1)}+W_R^{(2)}][O_R+O_R^2+O_R^3+...] \nonumber\\
     \times[1+R_1+R_2+...],
\eea
where the second term in right hand side represent possible multiple overhang and the third term, the possibility of
right steps after the overhang. $U_R$ may be rewritten as
\be
\label{right_overhang}
U_R=\bigg[W_R^{(1)}+W_R^{(2)}\bigg]\bigg[\frac{O_R}{1-O_R}\bigg]\bigg[1+\widetilde{R}\bigg],
\ee
where $\widetilde{R}$ is defined in Eq.~(\ref{right_left_wt}).

\subsubsection{Left Overhangs}

\begin{figure}
\includegraphics[width=\columnwidth]{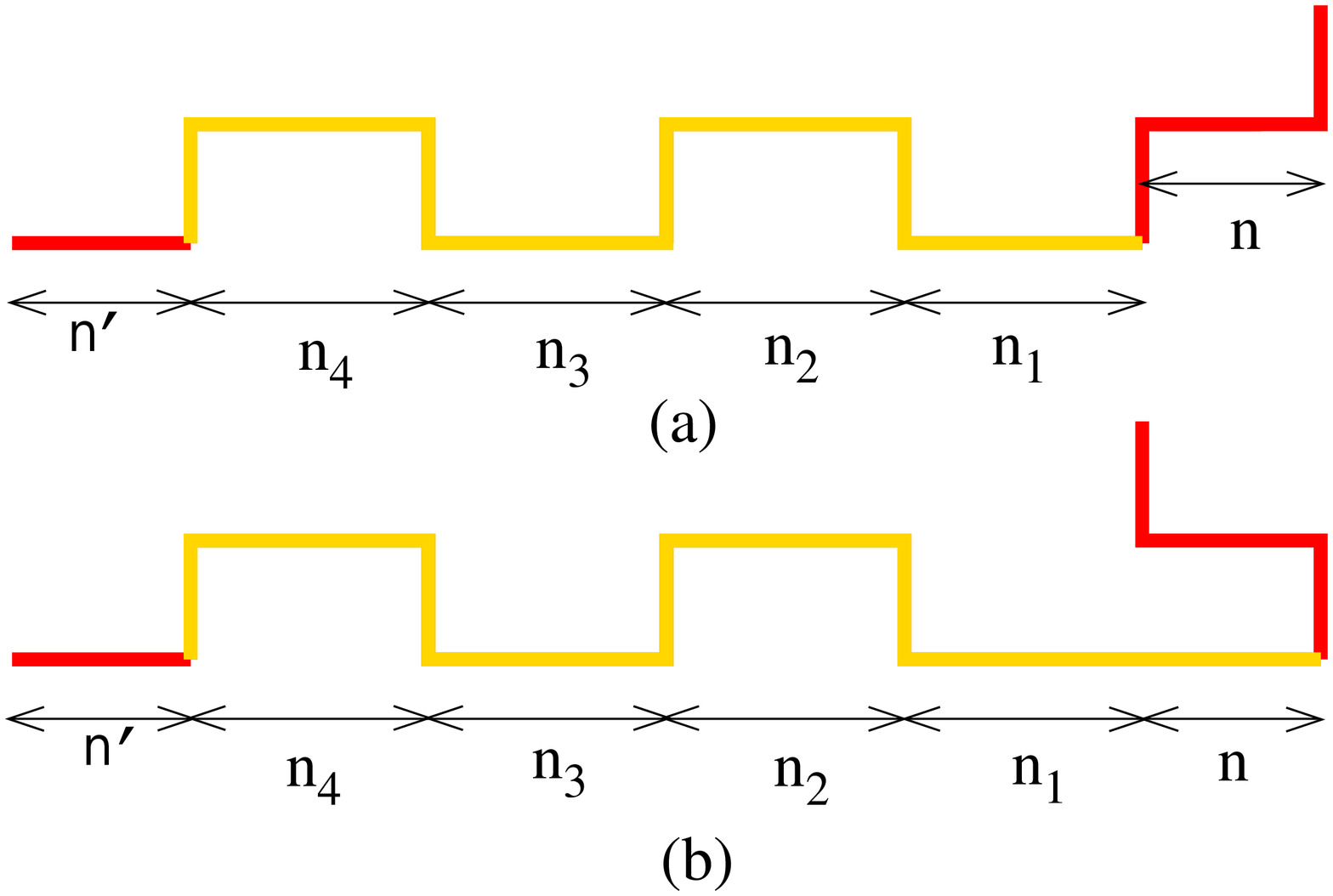}
\caption{Schematic diagram of two ways of taking two downward steps before the beginning  of an left overhang.
The first downward step may be followed by (a) left or (b) right step. Overhangs are denoted by yellow line.}
\label{f200}
\end{figure}

The weight for left overhangs may be calculated in a manner similar to that for right overhangs. 
There are two kinds of left overhangs depending on whether the first downward step is followed by leftward steps
[see Fig.~\ref{f200} (a)] or rightward steps [see Fig.~\ref{f200} (b)]. We split each of these into three parts, 
where the initial red line represents the first  two downward steps and the intervening region, the middle yellow line 
represents overhangs and final red line
represents leftward steps.
We denote these parts symbolically  by $\mathbb{W}_L^{(1)}(n)$, $\mathbb{W}_L^{(2)}(n)$, $\mathbb{O}_L(n_1,n_2)$ and ${\mathbb{L}}^\prime(n^\prime)$ respectively, where the 
superscript (1) and (2) refer to the two kinds of left overhangs,
and subscript $L$ stands for left. Thus, a generic left overhang of first kind [see Fig.~\ref{f200} (a)] is represented by
\be
\mathbb{W}_L^{(1)}(n)\mathbb{O}_L(n_1,n_2)\mathbb{O}_L(n_3,n_4)...{\mathbb{L}}^\prime(n^\prime),
\ee
and that of the second kind [see Fig.~\ref{f200} (b)] is represented by
\be
\mathbb{W}_L^{(2)}(n)\mathbb{O}_L(n+n_1,n_2)\mathbb{O}_L(n_3,n_4)...{\mathbb{L}}^\prime(n^\prime).
\ee

We now determine the  weights of the different parts constituting the left overhangs.
For the overhang of first kind, the total weight associated with the initial red part $\mathbb{W}_L^{(1)}(n)$ may be written as
\be
W_L^{(1)}=\sum_{n=1}^\infty wt[\mathbb{W}_L^{(1)}(n)]=D^2\sum_{n=1}^\infty L_{\Delta}=D^2\widetilde{L},
\ee
where $L_{\Delta}$ and $\widetilde{L}$ are given respectively in Eqs.~(\ref{left_weight}) and (\ref{right_left_wt}).
The weight $O_L$ associated with the left overhang is identical to that of the right overhang of similar shape, i.e.,
\be
O_L=\sum_{n_1}\sum_{n_2}wt[\mathbb{O}_L(n_1,n_2)]=O_R,
\ee
where $O_R$ is given in Eq.~(\ref{over_r}). 

 Now, consider the left overhangs of the second kind [see Fig.~\ref{f200} (b)].
The total weight associated with the first two downward steps and the first overhang may be written as
\bea
\mathcal{W}_L^{(2)}&=&\sum_n\sum_{n_1}\sum_{n_2} wt[\mathbb{W}_L^{(2)}(n)\mathbb{O}_L(n+n_1,n_2)] \nonumber\\
&=&\sum_{n=0}^\infty\sum_{n_1=1}^\infty D^2L_n\omega(n+n_1)\bigg[z_v\Omega(n+n_1-1,0)\nonumber\\
&+&z_s\Omega(n+n_1-2,0)\bigg]\lambda^{-3(n+n_1)/2}\sqrt{O_L},
\label{wl2}
\eea
where the summation over $n_2$ contributes $\sqrt{O_L}$. It is convenient to change the variable to $m=n+n_1$. In this new
variable Eq.~(\ref{wl2}) may be rewritten as
\bea
\mathcal{W}_L^{(2)}&=&\sum_{m=1}^\infty\sum_{n=0}^{m-1}D^2L_n m \omega(m) \bigg[z_v\Omega(m-1,0)\nonumber\\
&+&z_s\Omega(m-2,0)\bigg]\lambda^{-3m/2}\sqrt{O_L}.
\label{wl2s}
\eea
We define the functions
\be
\mathcal{F}_i(y)=y^{1+i}\frac{d}{dy^\prime}\bigg[G(y^\prime,0)\bigg]_{y^\prime=y}+iy^iG(y,0),
\ee
where $i=1, 2$ and $G(y,0)$ is the generating function as determined in Eq.~\eqref{eq gya}.
Doing the sums in Eq.~(\ref{wl2s}), we obtain
\bea
\mathcal{W}_L^{(2)}&=\frac{D^2}{a(0)}\bigg[\bigg(\frac{p_+x_+\sqrt{\lambda}}{x_+\sqrt{\lambda}-1}+\frac{p_-x_-\sqrt{\lambda}}{x_-\sqrt{\lambda}-1}\bigg)\mathcal{J}\bigg(\frac{1}{x_+\lambda^{3/2}},\frac{1}{x_-\lambda^{3/2}}\bigg)\nonumber\\
&-\frac{p_+x_+\sqrt{\lambda}}{x_+\sqrt{\lambda}-1}\mathcal{J}\bigg(\frac{1}{x_+^2\lambda^{2}},\frac{1}{x_+x_-\lambda^{2}}\bigg)\nonumber\\
&-\frac{p_-x_-\sqrt{\lambda}}{x_-\sqrt{\lambda}-1}\mathcal{J}\bigg(\frac{1}{x_+x_-\lambda^{2}},\frac{1}{x_-^2\lambda^{2}}\bigg)\bigg]\sqrt{O_L},
\eea
where
\bea
\mathcal{J}(x_1,x_2)&=&z_s\bigg[b_+\mathcal{F}_2(x_1)+b_-\mathcal{F}_2(x_2)\bigg]\nonumber\\
&+&z_v\bigg[b_+\mathcal{F}_1(x_1)+b_-\mathcal{F}_1(x_2)\bigg].
\eea
with $b_\pm$ as defined in Eq.~\eqref{eq:bplus}.

Finally, we calculate the total weight
associated with the final set of leftward steps that may be taken after the overhangs: 
\bea
\widetilde{L}^\prime&=&\sum_{n^\prime=0}^\infty wt[{\mathbb{L}}^\prime(n^\prime)]=\sum_{n^\prime=0}^\infty\omega(n^\prime)\lambda^{-n^\prime/2}\nonumber\\
&=&\bigg(\frac{b_+x_+\sqrt{\lambda}}{x_+\sqrt{\lambda}-1}+\frac{b_-x_-\sqrt{\lambda}}{x_-\sqrt{\lambda}-1}\bigg).
\eea
 The total weight of all walks with left overhang may now be computed. Let $\mathbb{U}_L$ represent all possible walks with at
least one left overhang. Let the total weight associated with these walks be 
$U_L$. Then we obtain
\bea
U_L=[W_L^{(1)}O_L+\mathcal{W}_L^{(2)}][1+O_L+O_L^2+...][\widetilde{L}^\prime],
\eea
where the second term in the  right hand side represents  multiple overhangs and the third term accounts for the
leftward steps after the final overhang.
Rewriting,
\be
\label{left_overhang}
U_L=\bigg[W_L^{(1)}O_L+\mathcal{W}_L^{(2)}\bigg]\bigg[\frac{1}{1-O_L}\bigg]\bigg[\widetilde{L}^\prime\bigg].
\ee

\subsubsection{Phase boundary}

 The formal generating function of all possible walks including those with overhangs may be written as
\be
\label{g_ov}
\mathbb{G}_{ov}=\frac{1}{1-(\mathbb{D}+\mathbb{D}\widetilde{\mathbb{R}}+\mathbb{D}\widetilde{\mathbb{L}}+\mathbb{U}_R+\mathbb{U}_L)}.
\ee
For every term in the expansion, one may uniquely identify a walk from top to bottom.
The weighted generating function corresponding  to $\mathbb{G}_{ov}$ may be written as
\be
\mathcal{G}_{ov}=\frac{1}{1-(D+D\widetilde{R}+D\widetilde{L}+U_R+U_L)}
\ee
As discussed earlier [see discussion following Eq.~(\ref{eq:G})], the generating function $\mathcal{G}_{ov}$ diverges at the
transition point when the interfacial tension $\sigma$ vanishes, and this condition is equivalent to

\be
\label{trans eq}
D+D\widetilde{R}+D\widetilde{L}+U_R+U_L=1,
\ee
where $D$, $\widetilde{R}$, $\widetilde{L}$, $U_R$ and $U_R$ are given in Eqs.~(\ref{down_weight}), (\ref{right_left_wt}), (\ref{right_overhang}) and
(\ref{left_overhang}) respectively.

In Fig.~\ref{f6}, blue line represents the critical line for activity with overhang.
It shows transitions at $z_s^{1/4}=0.733$ along SV line and at
$z_s^{1/4}=0.642$ along SD line. The density plot of critical line with overhang is shown in the 
Fig.~\ref{f7} by blue line. It shows transitions at $\rho_s=0.934$ along SV line and at
$\rho_s=0.779$ along SD line.

\subsection{Monte Carlo simulations \label{sec:numerical}}

In this section we determine the phase boundary numerically using grand canonical Monte Carlo simulations. Conventional
algorithms that include only local evaporation and deposition moves often do not equilibrate the system, within available
computer time, at high densities due to long-lived metastable states. Here, we implement an algorithm that 
updates two rows at a time using a transfer matrix based Monte Carlo algorithm~\cite{kabir2}. The algorithm
not only succeeds in equilibrating the system at high densities, but also at full packing.
This algorithm is a  generalization of a cluster algorithm that is able to equilibrate systems  of particles with large
excluded volume at high densities~\cite{joyjit1,kundu}.  In this Monte Carlo scheme~\cite{kabir2}, all particles that are  fully contained in
a $2 \times L$ track, consisting of two adjacent rows or columns of length $L$, are evaporated. The track is refilled with particles according to the correct weights 
in the partition function for the track,  subject to the constraints induced by particles  protruding into the track. The calculation of the restricted partition
function is done using standard transfer matrix technique, details of which may be found in the Supplemental Information of Ref.~\cite{kabir2}.

\begin{figure}
\includegraphics[width=\columnwidth]{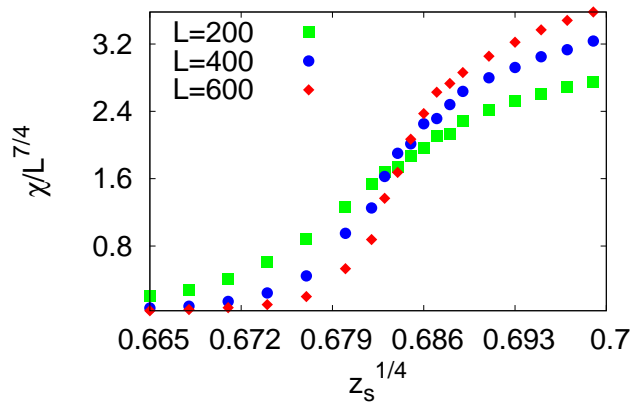}
\caption{Variation of  of $\chi/L^{7/4}$ with activity of squares $z_s^{1/4}$
for fixed activity of dimer $z_d=0.031$ for different system sizes. The curves cross at 
$z_{c}^{1/4}\approx 0.684$. Critical value of $z_0$
may be calculated from Eq.~(\ref{zsdv}).}
\label{crossing}
\end{figure}

The order parameter $Q$ is defined as
\be
Q=\sqrt{(\rho_{er}-\rho_{or})^2+(\rho_{ec}-\rho_{oc})^2},
\ee
where $\rho_{er}$, $\rho_{ec}$, $\rho_{or}$ and $\rho_{oc}$ are the   densities of heads of particles (including both dimers
and squares)
in even rows, odd rows, even columns and odd columns respectively.
Consider the second moment of the order parameter $Q$, defined as
\be
\label{chi2}
\chi=L^2\langle Q^2\rangle,
\ee
where $L \times L$ is the system size. Near the transition point $\chi$ obeys following
scaling law
\be
\label{scaling}
\chi\simeq L^{\gamma/\nu}f(\epsilon L^{1/\nu}),
\ee
where $\epsilon$ is the deviation from critical point
$\epsilon=z_s-z_c$,
and $f$ is the finite size
scaling function, and $\gamma$ and $\nu$ are the usual critical exponents. 
Since the model belongs to Ashkin-Teller universality class, one knows that
the exponent $\gamma/\nu=7/4$~\cite{kabir2}, independent of the parameters. 
From Eq.~(\ref{scaling}), we see that $\chi/L^{7/4}$ for different $L$ should cross
at  $\epsilon=0$, allowing the critical point to be estimated. An example is shown in
Fig.~\ref{crossing}. 
We fix the activity of dimer $z_d$ and vary the activity of square $z_s$ to get its critical value $z_c$.
Critical value of $z_0$ may be obtained from the normalization condition given in Eq.~(\ref{zsdv}). The data of full
phase boundary for activity plane and density plane are shown by blue dots in Fig.~\ref{f6} and Fig.~\ref{f7} respectively.

\section{\label{sec:5}Discussion}

To summarize, we calculated the interfacial tension between two differently ordered phases in a mixture of hard squares and
dimers, within two approximation schemes.  The estimates for the phase boundary between the ordered columnar phase
and disordered fluid-like or power law correlated phase was obtained by setting the interfacial tension to zero.
In the first calculation, we modeled the interface as having no overhangs. The estimates were improved by extending the
calculations to an interface where overhangs of height one were allowed. 
The estimates that we obtain for critical parameters are shown in Fig.~\ref{f6} and \ref{f7}, and are in good agreement with 
results from Monte Carlo simulations. The deviation from the numerical results are largest along the fully packed 
square-dimer (SD) line.
Along this line, we obtain the critical parameters  to be
$z_s^{1/4}=0.616$,   $\rho_s=0.714$ for interfaces without overhang,  and $z_s^{1/4}=0.642$, $\rho_s=0.779$
for interfaces with overhangs. These are to be compared with results from Monte Carlo simulations: $z_s^{1/4}=0.692$ and 
$\rho_s=0.843$. Along the square vacancy (SV) line, the calculation reproduces the results in Ref.~\cite{trisha2} for interfaces
without overhangs, but corrects the result for interfaces with overhangs. For the latter case, it was erroneously assumed in
Ref.~\cite{trisha2} that the contributions from left and right interfaces are identical.

In our calculations we assumed that the ordered phases have perfect order, thereby ignoring the presence of defects
in the bulk phases. Defects may be included in a systematic manner, as was done for the case of 
$2 \times d$ rectangles \cite{trisha2}. 
However, it was found that the corrections appearing from including overhangs were more dominant than 
that arising from including defects when $d$ was small as is the case for hard squares. At the same time, the calculations
for including defects is  involved and including the effect of two defects appears a formidable task. We have, therefore, ignored
the corrections due to the presence of defects.

The calculations could also be improved by taking into account multiple defects using the pairwise approximation~\cite{dipanjan}. 
Such an approach could improve the estimates for the phase boundary along the fully packed square-dimer line. This is a promising 
area for future study.

It will also be interesting to study mixtures of $m\times m$ squares and $m\times 1$ rods, $m=2$ being the square-dimer
problem.  Along the fully packed line, it would be possible to map the configurations to a vector height field. The 
ordered phase has a $2m$ fold symmetry. This will possibly change the nature of the transition from the ordered to
disordered phases. Also, along the fully packed line, it raises the possibility of an intermediate hexatic phase. For $m>2$,
the correlations even the fully packed $m$-mer problem is not known~\cite{ghosh2007random}. Thus, $m=3$ system (trimers+squares) would be a good starting point.

\begin{acknowledgments}
We thank Trisha Nath for helpful discussions.
\end{acknowledgments}


\end{document}